\documentclass[preprint,3p,times]{elsarticle}
\usepackage{amssymb}
\usepackage{amsmath}
\biboptions{square}
\usepackage{aas_macros}
\journal{Astroparticle Physics}

\begin{document}

\begin{frontmatter}
\title{Cherenkov $\tau$ Shower Earth-Skimming Method \\
	for PeV-EeV~$\nu_{\tau}$ Observation with Ashra}
\author[icrr]{Y. Asaoka}
\author[icrr]{M. Sasaki}
\address[icrr]{Institute for Cosmic Ray Research, University of Tokyo, \\
		Kashiwa, Chiba 277-8582, Japan}

\begin{abstract}
We describe a method of observation for PeV--EeV tau neutrinos using Cherenkov 
light from the air showers of decayed taus produced by tau neutrino interactions in the Earth. 
Aiming for the realization of neutrino astronomy utilizing the Earth-skimming tau neutrino detection technique,
highly precise determination of arrival direction is key due to the following issues: 
(1) Clear identification of neutrinos by identifying those vertices originating 
within the Earth's surface;
(2) Identification of very high energy neutrino sources.
The Ashra detector uses newly developed light collectors 
which realize both a 42 degree-diameter field-of-view and arcminute resolution. 
Therefore, it has superior angular resolution for imaging Cherenkov air showers. 
In this paper, we estimate the sensitivity of and cosmic-ray background resulting
from application of the Ashra-1 Cherenkov tau shower observation method. Both
data from a commissioning run and a long-term observation (with fully equipped
trigger system and one light collector) are presented.
Our estimates are based on a detailed
Monte Carlo simulation which describes all relevant shower processes 
from neutrino interaction to Cherenkov photon detection produced by tau air showers.  
In addition, the potential to determine the arrival direction of Cherenkov showers
is evaluated by using the maximum likelihood method. 
We conclude that the Ashra-1 detector 
is a unique probe into detection of very high energy neutrinos and their accelerators.
\end{abstract}

\end{frontmatter}

\section{Introduction}\label{sec:intro} 
Gamma-ray bursts (GRBs), important high energy transient phenomena, 
are plausible candidates for cosmic ray sources up to and including the highest energy regions 
\cite{WB95,Vietri95,Milgrom95}. 
Following the identification of GRB970228 with arcminute resolution by Beppo-Sax satellite 
\cite{BeppoSax97},
understanding of GRBs progressed dramatically, owing to the collaboration of satellite-supplied 
GRB triggers and follow up multi-wavelength observations.
As a result, the GRB standard model based on the particle acceleration in the internal/external shocks 
was established \cite{Rees-Meszaros92,Sari-Piran95,Piran99,Meszaros06}. 
On the other hand, observations of GRB and its afterglow made by Swift \cite{Gehrels04}
and Fermi \cite{Abdo09a} satellites revealed various phenomena which are difficult to explain 
in the framework of the standard model.
To resolve the various complicated aspects of the GRB physics mechanism, 
we would need ``multi-particle astronomy'' \cite{Sasaki00,Barwick00} 
which uses other particles in addition to photons.
In particular, neutrinos are the most important because 
with them it is possible to probe the optically thick region for electromagnetic components. 
As we can learn from Beppo-Sax's success,
identifying the source with superior angular resolution 
enables us to approach the physics mechanism by combining a number of 
observational results. 
To realize ``multi-particle astronomy'', arcminute resolution would be desirable
in the field of very high energy (VHE) neutrino observation. 

The All-sky Survey High Resolution Air-shower detector (Ashra) 
is a project which primarily aims to observe Cherenkov and
fluorescence lights from the lateral and longitudinal developments of 
very-high-energy (VHE) cosmic-ray air showers in the atmosphere. 
It uses newly developed light collectors (LCs) which realize both a 
42 degree-diameter field-of-view (FOV) and arcminute resolution. 
In particular, it can capture air-shower images with unprecedented precision. 
It is expected to determine the arrival direction of parent particle with high accuracy. 

\begin{figure}[hbt!]
   \begin{center}
    \includegraphics[width=0.6\hsize]{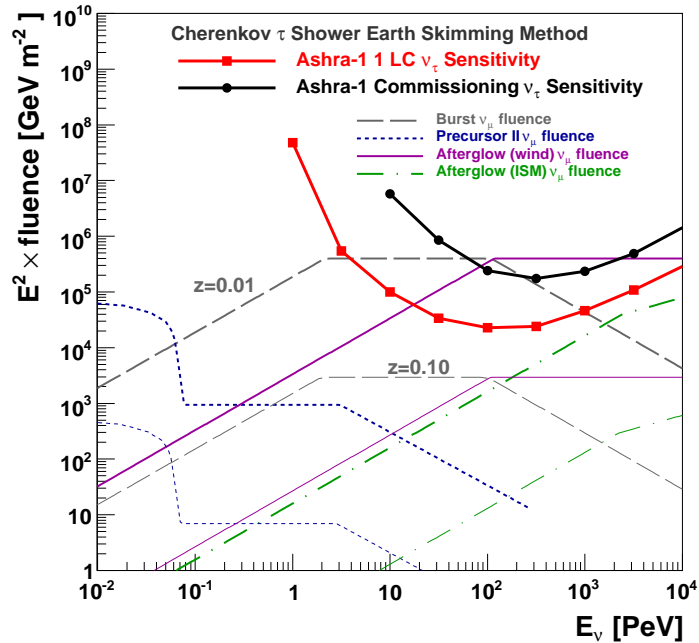}
   \end{center}
 \caption{The sensitivity of Cherenkov $\tau$ shower ES method
	with Ashra-1 where occurrence of GRB behind the mountain is assumed
	(see the text).
	Solid square shows the differential sensitivity of the Ashra-1 
	observation using one light collector and solid circle
	shows that of the commissioning observation.
	Note that differential sensitivity represents the fluence where 2.3 events
	are expected at the corresponding energy decade.
	\label{fig:flux}
	Also shown are the scaled neutrino fluence expected from GRB030329 occurred  
	at $z=0.01$ and $z=0.1$ in each phase of GRB \cite{RMW04}.
	The nearest and second nearest GRBs to date 
	were occurred at z=0.0085 (GRB980415 \cite{IAUC6895}) and z=0.0331 
	(GRB060218 \cite{GCN4792}), respectively.
	The dashed, dotted, solid and dash-dotted lines show the components of 
	burst, precursor, afterglow in wind environment, and afterglow in ISM environment, 	
	respectively.
	}
\end{figure}

As to detection methods for VHE neutrinos,  
there are methods which use water and/or ice as the neutrino target \cite{IceCubeGRBs,IceCubeGRB},
and those which utilize air showers, i.e., the 
deeply penetrating neutrino air-shower detection technique and 
recently proposed ``Earth skimming tau-neutrino ($\nu_{\tau}$) technique'' 
\cite{Domokos98,Letessier00,Athar00,Fargion02,Feng02}.
Some results have already been published with the Earth-skimming method 
(for example, see \cite{Abraham08}).
On the other hand, by combining Cherenkov emission in the optical band and 
``Earth skimming $\nu_{\tau}$ technique'' (Cherenkov tau shower Earth-skimming method; 
hereafter referred to as Cherenkov $\tau$ shower ES method),  
we can effectively survey an energy range which is hard to reach 
with the above mentioned technique. It is possible to achieve a maximum of sensitivity 
around 10--100~PeV in which the VHE neutrino signals from GRBs are expected.
Figure~\ref{fig:flux} shows the expected neutrino fluence from nearby GRB,
in which the calculated fluence from GRB030329 shown in Ref. \cite{RMW04} is scaled to
different redshifts($z$). The expected sensitivities of the Ashra-1 detector by using 
Cherenkov $\tau$ shower ES method are also plotted in the figure,
where occurrence of GRB behind the mountain is assumed.
The sensitivity calculation is performed by using the Monte Carlo simulation described in detail
in this paper.

This paper is organized as follows: Section~\ref{sec:ashra} introduces the Ashra-1 detector,
Section~\ref{sec:skim} describes the detailed method to simulate Earth-skimming $\nu_{\tau}$
events, including the discussion about the deflection during their propagation.
Section~\ref{sec:sen} shows the result of a sensitivity estimation. Section~\ref{sec:disc}  
discusses the systematic errors due to the incorporated Monte Carlo simulation, 
background estimation and angular reconstruction accuracy of $\tau$ shower events.
We conclude in Section~\ref{sec:concl}.

\section{Ashra-1 Detector}\label{sec:ashra} 
The All-sky Survey High Resolution Air-shower detector Phase I (Ashra-1) is
an optical-telescope based detector system \cite{Sasaki08} optimized to detect VHE particles. 
Ashra-1 is distinguished by two features:   
(1) an ultra wide optical system in which 
42-degree FOV is demagnified to 1 inch by using 
photon and electron optics \cite{PLI11}; 
(2) the high resolution imaging system with trigger. 
Ashra-1 combines these unique features, resulting in very cost-effective pixels 
compared to conventional photomultiplier arrays at the focal
surface of an optical telescope. 
Ashra-1 can observe the whole sky with arcminute resolution  
with 12 detector units pointing at different directions,  
where a detector unit consists of a few LCs pointing at the same direction. 

The Ashra-1 detector system is designed so that the focal image is split into 
trigger/image capture devices after amplification.
This feature enables us to simultaneously obtain 3 kind of phenomena which have
different time scales, i.e., Cherenkov emission (ns), fluorescence ($\mu$s), and starlight (s)
without sacrificing the signal to noise ratio.
By fully utilizing these distinct features, 
Ashra is aiming to undertake the full-fledged astronomical observation using VHE particles
and trying to achieve the detection of VHE neutrinos for the first time 
using the Earth and the nearby mountain as the 
target \cite{AshraCNeu}, 
an arrangement provided by the Ashra-1 observatory being
located on the Mauna Loa (3,300m) on Hawaii Island, opposite to the Mauna Kea. 
It can also be used to optically observe transient objects like GRBs 
as it monitors the whole sky simultaneously \cite{GCN8632,GCN11291}. 

\section{Earth Skimming Tau Neutrino}\label{sec:skim}
\subsection{Neutrino detection method}
\begin{figure}[hbt!]
   \begin{center}
    \includegraphics[width=0.6\hsize]{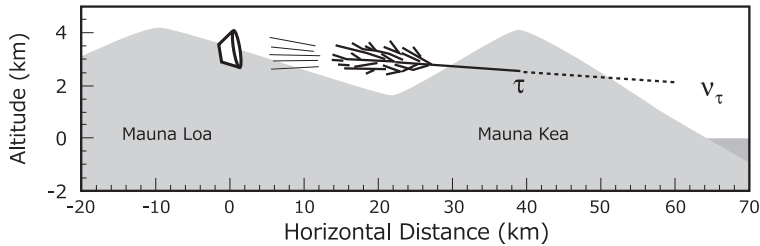}
   \end{center}
 \caption{
	Schematic view of Cherenkov $\tau$ shower ES method.\label{fig:mtskim}
	Mauna Kea is used as the target mass for neutrino charged current interaction.
	The produced air shower is observed from Mauna Loa.
	In addition to the fact that the mountain can be viewed with
	large solid angle from the observatory, the distance of about 30km from
	the observatory to the Mauna Kea surface is appropriate for the air shower
	development in 10--100~PeV energy range, 
	resulting in the huge advantage of the Ashra-1 observatory.}
\end{figure}

To detect VHE neutrinos, a large target volume is required in order to compensate for the 
very small neutrino-nucleon cross section. 
On that basis, the secondary particles produced by the first neutrino interaction 
are required to be detected through some method. 
The detection method using water and ice as a target detects Cherenkov light from
secondary muons taking advantage of the fact that ice and water are to some extent optically 
transparent. 
This method can be categorized by identical target and detection volume. 
On the other hand, the detection method using 
deeply penetrating air showers from higher-energy neutrinos
uses atmosphere as a target and detection volume.
This method enables achieving a huge detection volume as the atmosphere has 
very high transmittance. However, 
it is difficult to obtain a larger target mass due to low atmospheric density. 
The detection method called Earth-skimming $\nu_{\tau}$ technique
\cite{Domokos98,Letessier00,Athar00,Fargion02,Feng02}
can realize a huge target mass and detection volume at the same time 
by separating the target and detection volume 
utilizing the interaction process of $\nu_{\tau}$.
The detection method is described as follows (See Fig.~\ref{fig:mtskim}): 
a VHE $\nu_{\tau}$ interacts in the Earth or mountain and produces a tau lepton ($\tau$). 
$\tau$ penetrates the Earth and/or mountain and enters in the atmosphere. 
Subsequently, it decays and produces an air shower.
Cherenkov photons from the air shower are detected.
Owing to the separation of the first interaction where $\nu_{\tau}$ produces 
$\tau$ and $\tau$ decay generating air shower,
air shower observation becomes possible while preserving the huge target mass required  
to compensate for the low cross section of the first interaction.  
For this detection method, it is crucial for the $\tau$ to go through the Earth and/or   
mountain before its decay, and to develop the air-showers in the atmosphere in front 
of the detector after its decay. Note that the decay length of 100 PeV $\tau$ is 4.9km.
Here, ``Cherenkov $\tau$ shower ES method'' is defined as the detection 
method which detects Cherenkov photons from $\tau$ shower appearing from the Earth or 
the mountain fully utilizing this feature.
Mauna Kea is over 3,200 km$^3$ in volume and 9.3 tera tons in mass \cite{MaunaKea}, 
making it the perfect location to implement an experiment using this technique. 

\subsection{Deflection from parent tau neutrinos}\label{sec:tauang} 
This section describes deflection of Cherenkov $\tau$ shower compared to the 
arrival direction of parent $\nu_{\tau}$ 
in order to estimate the ability to trace back the direction 
of the detected air shower to the accelerator. 
We evaluated the deflection of the propagating particle in each step from the 
neutrino charged current interaction to $\tau$ propagation in the Earth,  
tau decay, and production of the extensive air shower.
We used PYTHIA~(ver.~6.1) \cite{PYTHIA6154} to evaluate the neutrino charged current interaction.
Since $P_t < M_W$ (where $M_W$ denotes the mass of the W boson which mediates the charged 
current interaction) and $P_t$ denotes the transverse momentum of the produced $\tau$,  
the deflection angle of $\tau$ ($\Delta \theta_{\tau}$) with respect 
to the parent $\nu_{\tau}$ should be less than 0.3~arcmin for $E_{\tau}>1$~PeV.
The simulation results with PYTHIA were consistent with this.  

Second, we used GEANT4~(ver.~9.3) \cite{Geant4} to evaluate the $\tau$'s deflection due 
to propagation in the Earth. 
To estimate the energy loss of high energy leptons, the following parametrization
is generally adopted \cite{Dutta2001}:
\[ -\left < \frac{dE}{dX} \right > = \alpha + \beta E, \]
where $\alpha$ denotes the nearly constant parameter determined by ionization loss,
and $\beta$ denotes the radiative energy loss due to Bremsstrahlung, pair production
and photonuclear interaction.
Since radiative energy loss is dominant for high energy $\tau$s, 
these high energy process must be included in the "Physics List" of GEANT4.
Thus, we applied the following processes originally defined for muons to $\tau$s,
and estimated the deflection after propagating through 10km rock. 
\begin{itemize}
\item G4MuBremsstrahlung: Bremsstrahlung
\item G4MuPairProduction: $e^+e^-$ pair production
	\footnote{we modified the original G4MuPairProduction so that
	the momentum preserves including the produced particles, resulting
	in the inclusion of deflection.}
\item G4MuNuclearInteraction: Photonuclear Interaction
\end{itemize}

\begin{figure*}[hbtp]
   \begin{center}
 \begin{tabular}{cc}
  \begin{minipage}{0.45\hsize}
   \begin{center}
    \includegraphics[width=0.8\hsize]{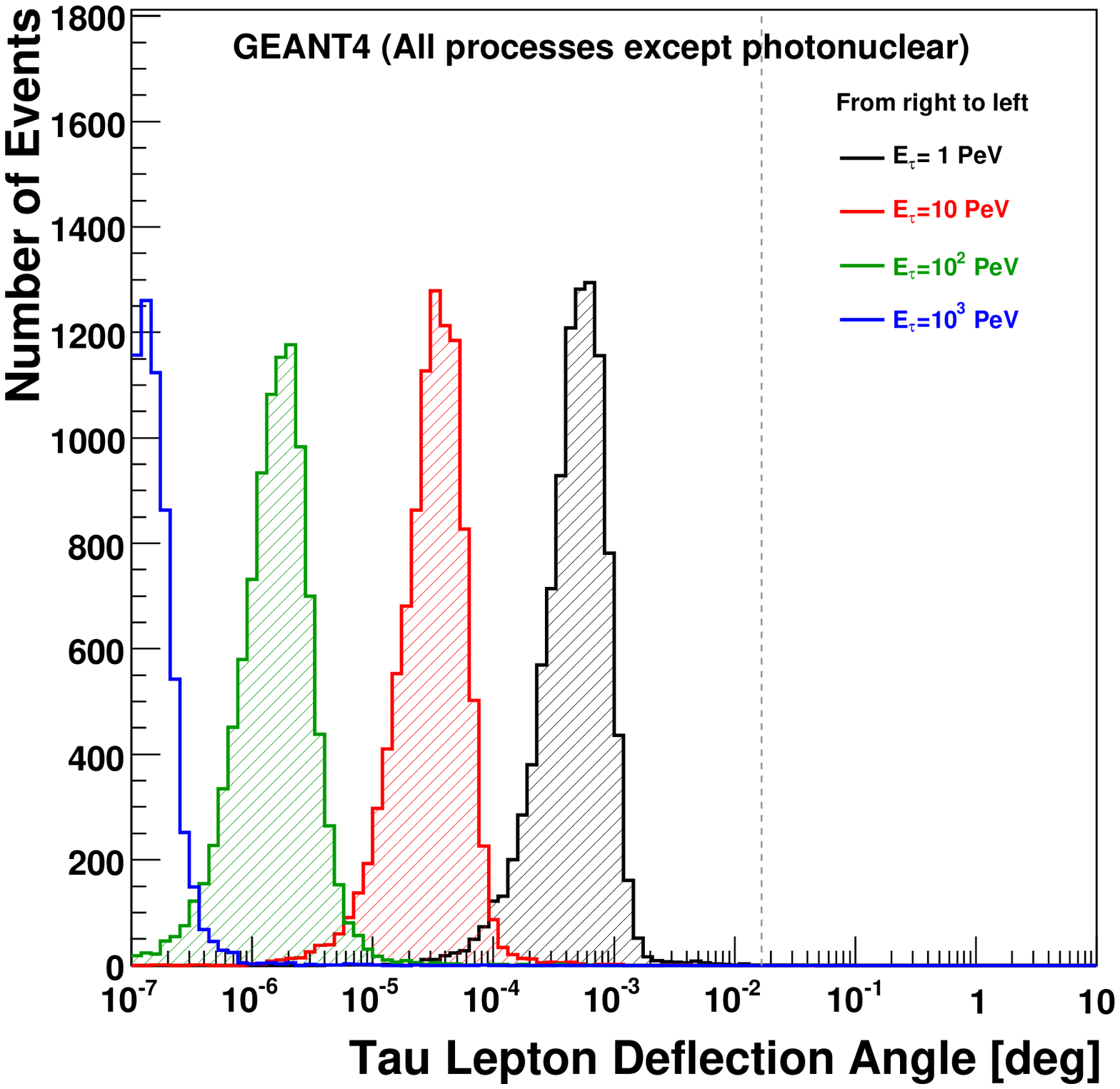}
   \end{center}
  \end{minipage}
  \begin{minipage}{0.45\hsize}
   \begin{center}
    \includegraphics[width=0.8\hsize]{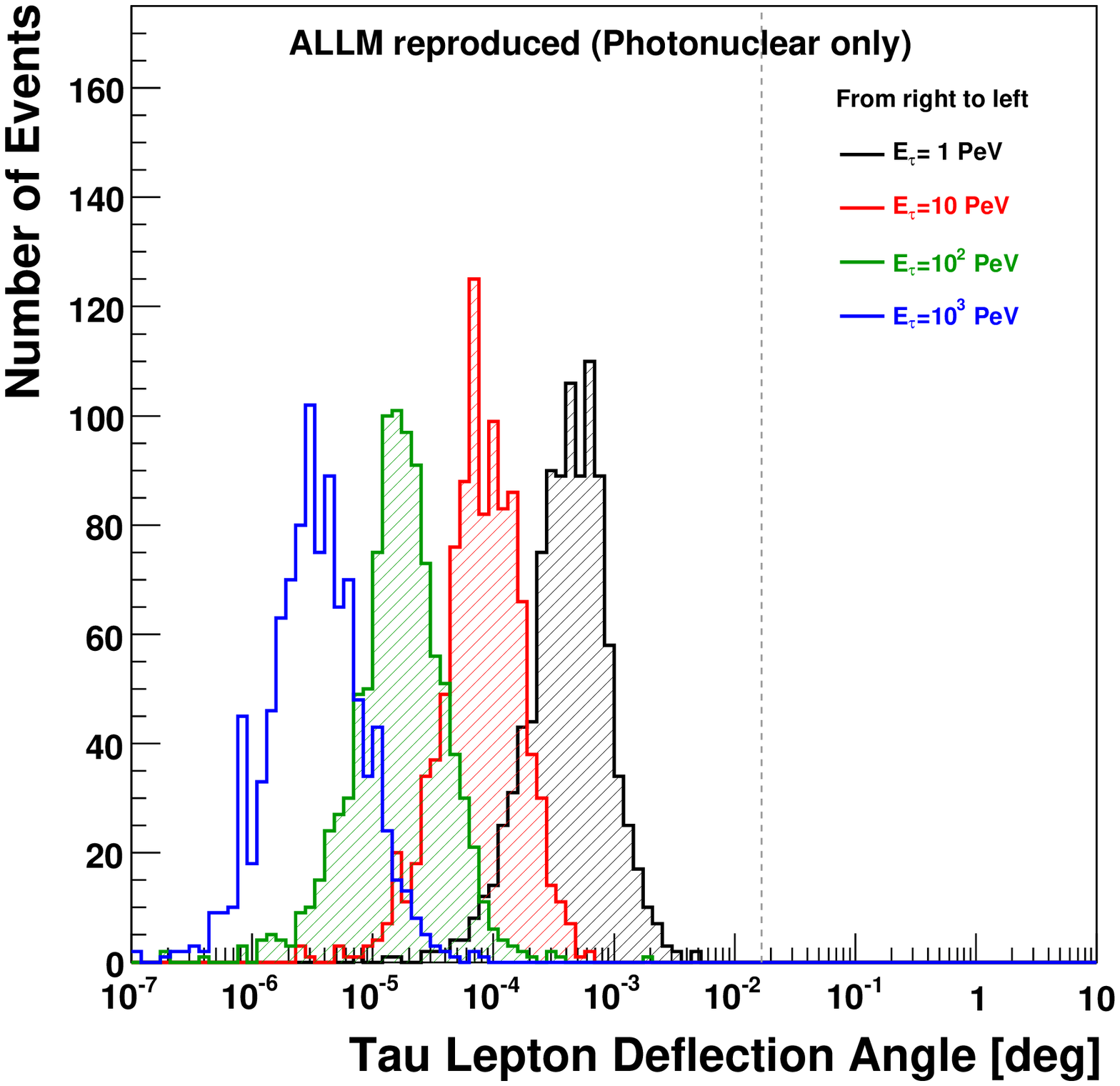}
   \end{center}
  \end{minipage}\\
 \end{tabular}
 \caption{The simulation results of deflection angle 
	of $\tau$s after propagating 10km of rock. 
	({\it Left}) the GEANT4 results including all the high energy processes
	except for photonuclear interaction.
	({\it Right}) the results of photonuclear interaction simulated 
	by handmade simulation.
	Note that the decay of $\tau$s was switched off for the above simulations.
	The hatched histograms indicate that the $\tau$ range is less than 10km.}
 \label{fig:defl}
 \end{center}
\end{figure*}

To validate our GEANT4 simulation, we compared the energy dependence of $\beta$
for Bremsstrahlung, pair production, and photonuclear interaction to Ref. 
\cite{Dutta2001}.
The $\beta$ energy dependence agreed well for the former two processes, 
but we found that GEANT4 produced smaller values for photonuclear interaction 
at higher energy and that the difference was 3 times at 10$^8$ GeV. 
Therefore we wrote a toy Monte Carlo simulation for photonuclear  
interaction using the formalism of Refs. \cite{ALLM, DiffCS}. 
In this simulation, the energy
dependence of $\beta$ in the relevant energy range 
was well reproduced within $\pm$1\% 
although the absolute value 
of $\beta$ was 30\% higher than that in the original work.
Note that our estimate on $\tau$'s deflection due to photonuclear interaction
can be assumed to be conservative since larger deflection is expected from 
higher $\beta$.
Figure~\ref{fig:defl} shows the simulation results of deflection angle 
of $\tau$s after propagating 10km of rock. 
Left panel shows the GEANT4 results including all the high energy processes
except for photonuclear interaction, and the right panel shows the 
results of photonuclear interaction simulated by our toy Monte Carlo simulation. 
These results indicate that photonuclear interaction becomes dominant for 
deflection at 1 PeV and higher.
Note that the decay of $\tau$s was switched off for above simulations and 
the hatched histograms indicate that the $\tau$ range is less than 10km.  
For example, the $\tau$ range is 4.9~km at 100~PeV. 
We conclude that the deflection angle of $\tau$s with energy greater
than 1~PeV is much less than 1 arcmin.

Next, the deflection due to $\tau$ decay is estimated by using the output of 
TAUOLA\cite{TAUOLA24} taking the $\tau$ polarization into account.  
From the mass of the $\tau$s, the deflection angle must be less than 1 arcmin 
if the energy of the secondary particle is higher than 13~TeV.
Using TAUOLA outputs in our simulation sample of PeV-energy neutrinos, 
we found that the probabilities to produce decay particle with its energy less 
than 13~TeV were 1.7\% and 0.038\% for parent neutrino energies of 1~PeV and 10~PeV,
respectively. Furthermore, none of the simulated events with $E_{\tau} > 13.7$~TeV 
satisfied the trigger requirements in both 
of Commissioning and Ashra-1 detector configurations. 
We conclude that the deflection angle between   
decay particles which produce the air shower and parent $\nu_{\tau}$ was less than 1 arcmin
for detectable events. 

Finally, the direction of the hadron air shower was evaluated using CORSIKA~(ver.~6.900). 
At the shower maximum, we compared the direction of the parent particle (charged pion)
to that of electrons and positrons 
which are the dominant producers of Cherenkov photons.
We found that on average the direction of electrons and positrons and the 
parent particle of the air shower was coincident within 0.1~$^\circ$ at 1~PeV.
In conclusion, we found that the arrival direction of PeV $\nu_{\tau}$s was 
preserved within 0.1~$^\circ$ including the hadron air-shower generation.
The accurate reconstruction of the arrival direction by means of fine imaging
will be a very powerful technique to determine point sources of PeV $\nu_{\tau}$s. 

\subsection{Monte Carlo simulation}
To investigate the neutrino identification capability and sensitivity,
it is vital to develop a detailed Monte Carlo simulation which simulates all processes from 
the neutrino interaction to detection of Cherenkov photons produced by decayed $\tau$ air showers. 
An earlier study mainly based on an analytic approach can be found in Ref.\cite{NuTel}.
In our simulation, 
the first step is to generate the $\nu_{\tau}$s from the certain area and solid angle which 
is large enough compared with our detector's detection volume. 
In this step, $\tau$ production due to the charged current interaction of $\nu_{\tau}$,
$\tau$ propagation in the Earth, and the determination of the decay point of $\tau$s emerging from 
the Earth are simulated.
These processes calculate the $\tau$ flux emerging from the Earth and subsequent decay  
in the atmosphere.
The flux is calculated for $\nu_{\tau}$s whose energies range 
from 10$^{15}$ eV (1~PeV) to 10$^{20}$ eV (100~EeV) in steps of 0.5 decades. 
The simulation code used in this step is based on the Earth skimming $\nu_{\tau}$ 
simulation code described in Ref.\cite{TANeu}.
The code is further developed to include the mountain skimming effect. 
Neutrino and $\tau$ interactions are updated from Ref.\cite{TANeu} in part 
in a relation with the previous work described in Ref.\cite{Noda09} 
and are described in the following.
The code is hereafter denoted as TauSim. 
Considering the direction and energy of $\tau$s emerging from the Earth and
removing the events with no probability of detection at this step, the efficiency
of the simulation was substantially improved.

For the neutrino-nucleon cross section, 
we used the calculation \cite{Gandhi98} 
based on CTEQ4-DIS parton distribution functions (PDFs).
Inelasticity is considered by fitting its energy dependence shown in Ref. \cite{Gandhi96}
to a quadratic expression.
To estimate the energy losses due to propagation of $\tau$s in the Earth,
we used the parametrization shown in Ref. \cite{Tseng03} based on Ref. \cite{Dutta2001}.
As the density of the Earth's crust, we took 2.9~g/cm$^{3}$ \cite{PREM}
since the main component of the Earth's crust around Hawaii is basalt. 

The second  step of our simulation deals with $\tau$ decay, air-shower generation, 
and detection in the detector.
Four vectors of secondary particles produced by $\tau$ decay were calculated by
TAUOLA version 2.4 \cite{TAUOLA24}.
The polarity of $\tau$ was set to $-1$ \cite{TANeu}.
The decay mode was determined for each $\tau$ obtained in the first step, resulting in 
1--4 secondary particle(s).
Then, an air shower was generated taking into account the geometrical relationship between the  
air shower and the detector. We used CORSIKA \cite{CORSIKA} to simulate the air showers. 
Since high energy air-shower generation requires a large amount of CPU power, we used 
a thinning algorithm. To keep the fineness of the shower image taken by the detector 
as much as possible, we adopted a smaller value of thinning level $= 10^{-7}$.  
We used QGSJET \cite{QGSJET01} as high-energy hadron interaction model.  
In CORSIKA, incoming photons to the detector's FOV were recorded 
by using the IACT package
\footnote{Since the IACT package in CORSIKA ver. 6.900 cannot receive the thinning 
information correctly, we fixed it to use in our simulation.},
and  used in the detector simulation.  
The detector simulation incorporated the light collection area obtained by 
ray tracing of the optical system, measured transmittance, and quantum efficiency of  
photoelectric lens image intensifier tube \cite{PLI11}.  
The detailed calibration method and final validation using cosmic ray  
observation in commissioning phase is described in Ref.\cite{AshraCNeu}. 
Trigger decision and acquisition of fine images 
were carried out in this step. Since CORSICA output includes the generation height and wavelength 
of detected Cherenkov photons, attenuation in the atmosphere was considered here.

\begin{figure}[hbt!]
   \begin{center}
    \includegraphics[width=0.65\hsize]{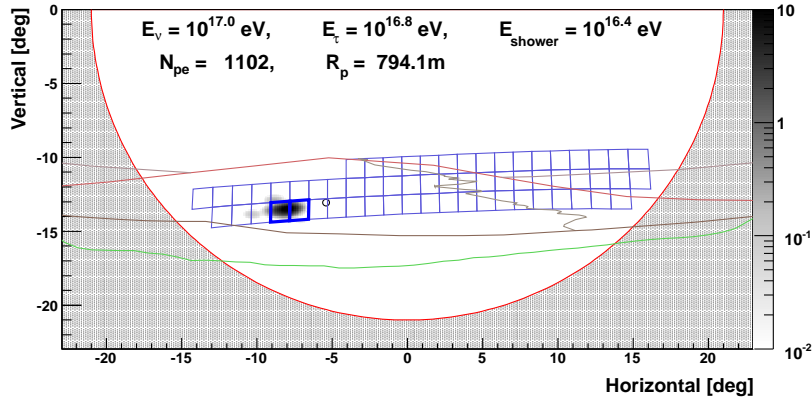}
   \end{center}
   \caption{\label{fig:evdisp} An example of the neutrino shower image 
	obtained by the detailed Monte Carlo simulation
	with the configuration of commissioning observation.
	The origin of horizontal and vertical axis 
	is set to the center of the light collector's FOV, 
	which is located at an altitude of 11.7$^\circ$  and an azimuthal 
	angle of 22.1$^\circ$.
	Boundary (large red circle) 
	between the inside (open circle) and outside (hatched area) of the FOV
	of the Ashra-1 light collector, which faces Mauna Kea,
	and the layout of trigger pixel FOVs 
	(blue boxes, firing trigger pixels with thick blue boxes) 
	for commissioning observation are shown.
	The ridge lines of Mauna Kea (red) and Mauna Loa (green) mountains, 
	the horizon, and Mauna Kea access road are also shown (gray).
	Open circle corresponds to the assumed typical point source coordinate 
	with altitude $-1.7^\circ$ and azimuthal angle $16.2^\circ$.
	} 
\end{figure}

Figure~\ref{fig:evdisp} shows an example of neutrino shower events 
simulated for the detector configuration of commissioning observation
obtained by the Monte Carlo simulation described above.
The origin of horizontal and vertical axis 
in Fig.\ref{fig:evdisp} is set to the center of the light collector's FOV, 
which is located at an altitude of 11.7$^\circ$  and an azimuthal 
angle of 22.1$^\circ$.
The unique capabilities of the Ashra detector enable us to obtain unmatched detailed images of 
Cherenkov showers. 
In the commissioning observation, we used a limited 62 channels of photo multipliers 
as trigger sensors to cover the view of the surface area of Mauna Kea as much as possible 
to maximize the sensitivity \cite{AshraCNeu}.
Adjacent-two logic was applied to trigger the fine imaging to reduce the background events
due to the fluctuation of night sky background. 
In the Ashra-1 observation with final configuration, the trigger pixel size will be halved 
(to one quarter of the current pixel area) and the entire FOV of the LC will 
be covered by the trigger sensor.

\section{Sensitivity}\label{sec:sen}
\begin{figure*}[hbtp]
   \begin{center}
 \begin{tabular}{cc}
  \begin{minipage}{0.45\hsize}
   \begin{center}
    \includegraphics[width=0.95\hsize]{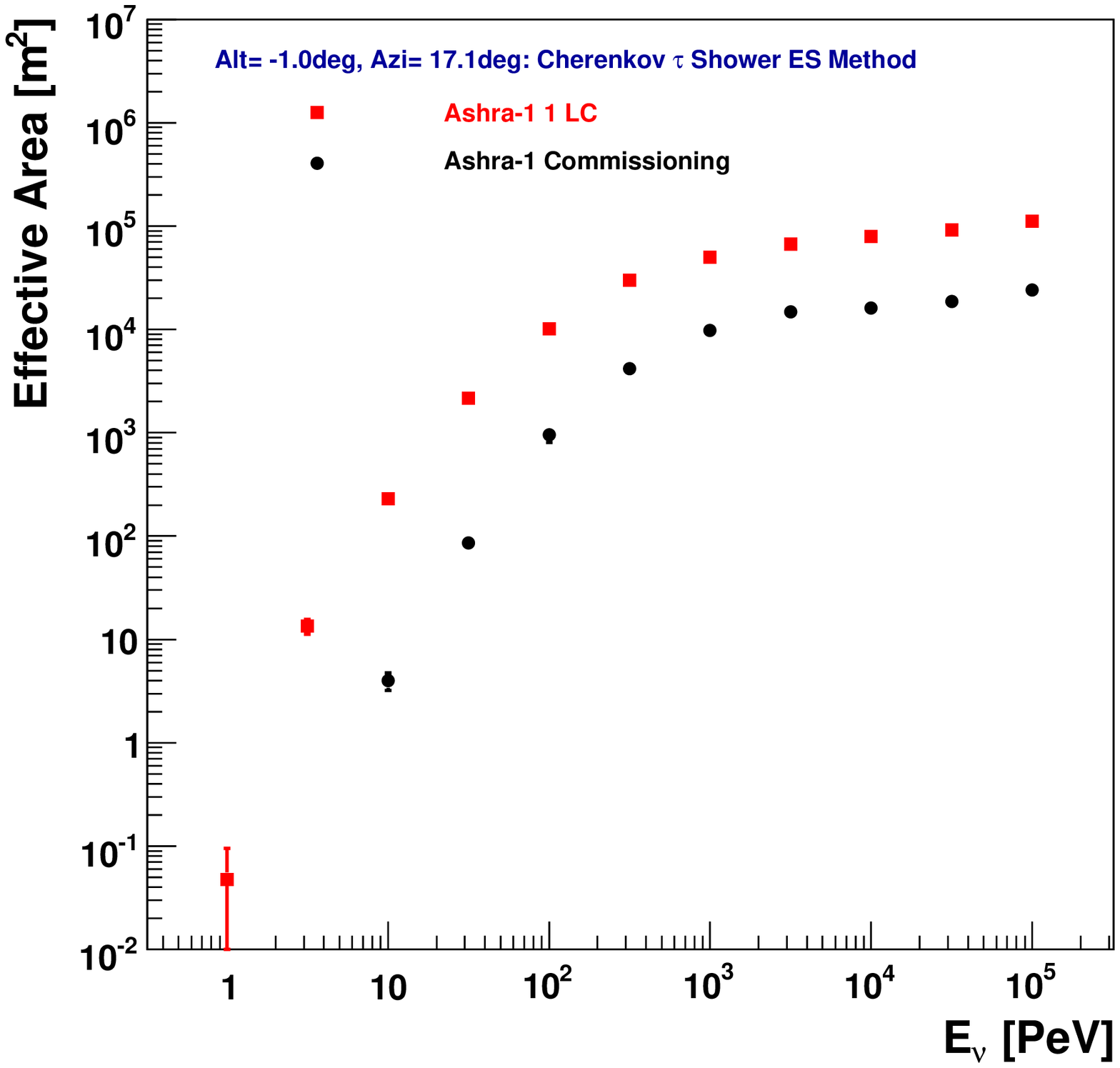}
   \end{center}
  \end{minipage} &
  \begin{minipage}{0.45\hsize}
   \begin{center}
    \includegraphics[width=0.95\hsize]{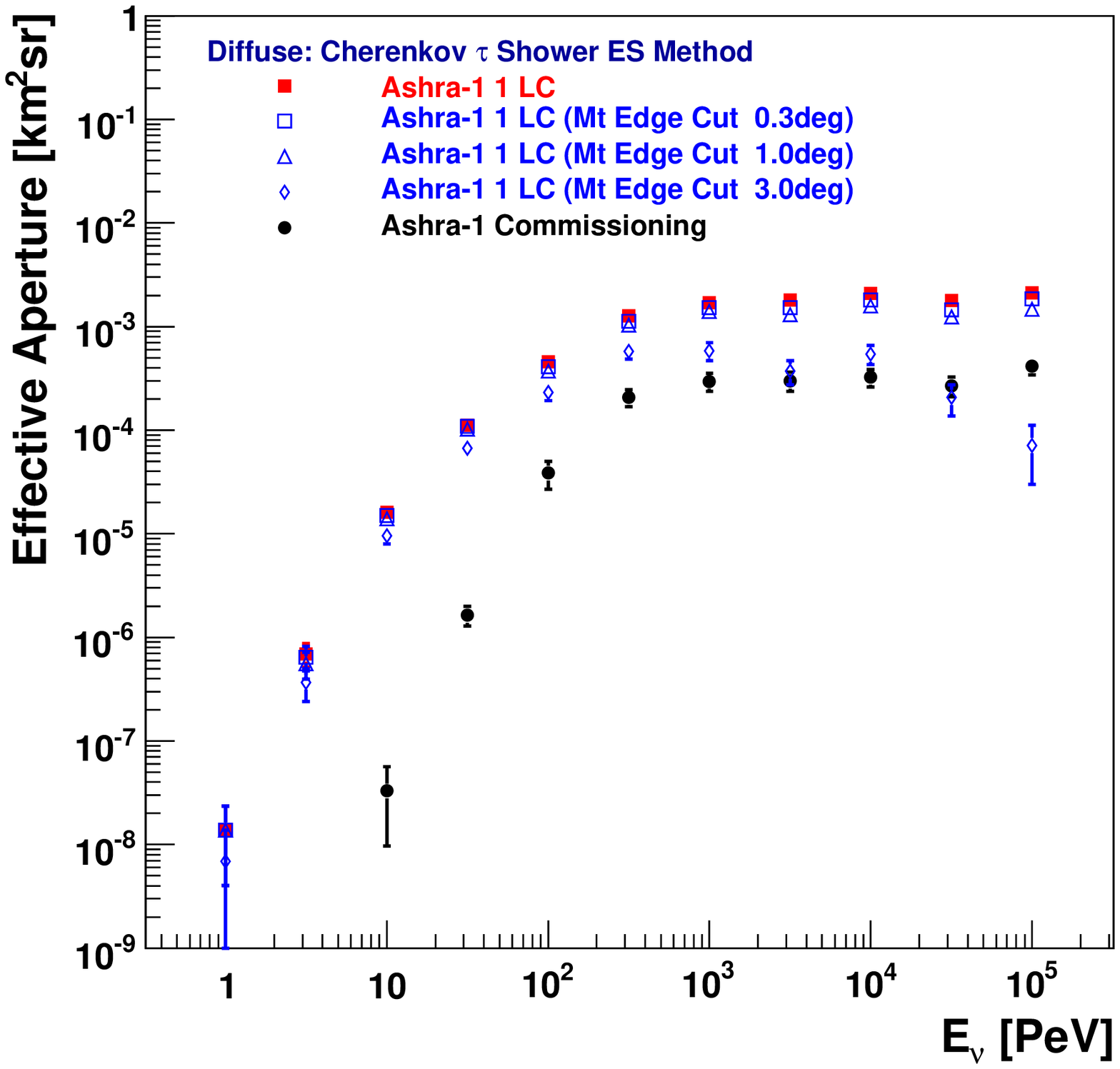}
   \end{center}
  \end{minipage}\\
 \end{tabular}
 \caption{ Typical effective area ({\it Left}) and effective aperture ({\it Right}) of 
	Cherenkov $\tau$ shower ES method with Ashra-1.
	The effective area was estimated in the case of an altitude of $-1.0^\circ$ 
	and an azimuthal angle of $17.1^\circ$ as an arrival direction of $\nu_{\tau}$s.
	Red solid squares and black solid circles represent the sensitivity of 
	1 light-collector observation and the commissioning observation, respectively.
	Blue open squares, blue open triangles, and blue open diamonds represent the
	effects of mountain-edge cut of 0.3$^\circ$, 1.0$^\circ$, and 3.0$^\circ$, 
	respectively. For explanation of mountain-edge cut, see the text.}
 \label{fig:aper}
 \end{center}
\end{figure*}

The effective area ($A_{\rm eff}$) is calculated for each $\nu_{\tau}$ energy
($E_{\nu_{\tau}}$) ranging from 1~PeV to 100~EeV in steps of 0.5 decades.
At first, from the direction of the point source,
$N_{\rm total}$ $\nu_{\tau}$ events are uniformly  
generated in the large area ($A_{\rm eff}^0$) containing Mauna Kea.
Assuming that $N_{\rm det}$ are finally triggered and detected from 
$N_{\rm total}$ events, the following equation gives $A_{\rm eff}(E_{\nu_{\tau}})$
as a function of neutrino energy.
\[ A_{\rm eff}(E_{\nu_{\tau}}) = A_{\rm eff}^0 
	\frac{N_{\rm det}(E_{\nu_{\tau}})}{N_{\rm total}(E_{\nu_{\tau}})}.	 \]
The left panel of Figure~\ref{fig:aper} shows an overlay of the effective area of 1-LC 
observation with final configuration and of the commissioning 
observation assuming typical point source coordinates with altitude $-1.0^\circ$ and
azimuthal angle $17.1^\circ$, which is indicated as the open circle in Fig.\ref{fig:evdisp}.
The commissioning observation used the measured trigger threshold described in Ref.\cite{AshraCNeu},
and 1-LC observation used the trigger threshold of 20 photoelectrons per pixel which corresponds 
to $\sim$1/7 of that in the commissioning observation. 
In both cases adjacent-2 trigger logic was used. 
The same neutrino shower samples were used to estimate the sensitivities for the detector 
configuration of commissioning and 1-LC observations although trigger simulations 
were carried out separately.
The right panel of Figure~\ref{fig:aper} shows the effective aperture
for a diffuse neutrino source where the neutrino direction and incident position are uniformly scanned.
In both cases of point and diffuse sources, improved sensitivity of 1-LC observation 
was achieved due to halved trigger pixel size, full coverage of light collector's FOV by trigger pixels,
and lower trigger threshold. Although there is no additional cut applied to estimate the effective area 
and effective aperture other than trigger decision, the triggered signal tends to have a much larger number of
photoelectrons than the minimum number required for a triggerable signal even in the 1-LC configuration. 
For example, the sensitivity weighted average of photoelectrons for triggered events 
in the 1-LC configuration was estimated to be $\sim$1000 for neutrino shower  
samples generated to estimate the diffuse neutrino aperture.  
Note that the sensitivity weight for each neutrino energy was calculated 
assuming a typical $E_{\nu}^{-2}$ flux. In the resultant photoelectron
distribution, $\sim$2\% of events have less than 100~photoelectrons.
The bunch size distribution of the generated showers in the CORSIKA output
was also studied and it was found that 99\% of them had 20 bunches or more 
(at least 10 times of trigger pixels to be fired) 
when detected number of photoelectrons was 40 or more, resulting in the validation 
of the applied thinning level of 10$^{-7}$ in terms of sensitivity estimation.

\section{Discussion}\label{sec:disc}
\subsection{Systematic errors}\label{syst}
The dominant systematic errors 
associated with the sensitivity of
Cherenkov $\tau$ shower ES method can be summarized as follows:
\begin{itemize}
\item neutrino-nucleon interaction cross section;
\item $\tau$ propagation in the Earth;
\item density of the Earth's crust and the geodetic model. 
\end{itemize}
In the following, the systematic uncertainties which worsen the resultant 
90\% upper limit on $\nu_{\tau}$ fluence and/or flux are considered.

At first, the systematic uncertainty of the neutrino-nucleon interaction cross section
was addressed. As explained, we used the calculation based on CTEQ4 \cite{Gandhi98}
for the total cross section of the neutrino-nucleon interaction. 
However, there is a large model uncertainty at higher energies due to 
inconsistency between PDF sets \cite{Anchordoqui06,Kutak03,Henley06,Sarkar08,Berger08}.
Among these models,
Refs. \cite{Anchordoqui06,Sarkar08} are based on the next-to-leading order linear dynamics 
[Dokshitzer-Gribov-Lipatov-Altarelli-Parisi (DGLAP)] constrained by HERA data.
The Auger experiment quoted an systematic uncertainty of $+5$\%$-9$\% for the expected neutrino rate 
\cite{AugerES09} due to cross section uncertainty based on Ref. \cite{Sarkar08}.
To include non-linear effects which become important at smaller Bjorken $x$, 
Ref. \cite{Kutak03} uses unified DGLAP/Balitsky-Fadin-Kuraev-Lipatov (BFKL) evolution 
supplemented by screening effects as well as nuclear shadowing. 
Ref. \cite{Henley06} uses color glass condensate (CGC) formalism which fits data from RHIC.
We took calculations based on a dipole model \cite{Kharzeev04} as the representative 
of CGC models, since they have the advantage that the BFKL anomalous dimension is built in \cite{Henley06}, 
and since the other models based on a different dipole model \cite{Bartels02} 
require a big assumption to determine the gluon distribution
function at smaller Bjorken $x$ \cite{Henley06}.
Ref. \cite{Berger08} considers the imposition of Froissart unitarity not to violate the Froissart bound
at higher energy.
From the largest difference between these models and the CTEQ4 parameterization in the relevant energy range, 
the systematic error of the neutrino-nucleon interaction cross section was estimated to be $-$35\%. 

Next, we evaluated the systematic errors of $\tau$
propagation in the Earth.
We used the approach from Refs.\cite{Dutta2001, Armesto08} 
to calculate the energy loss by the $\tau$ during 
propagation through the Earth and the mountain at PeV energies. 
The uncertainty due to the structure function used in the energy loss calculation
defined as the discrepancy 
between predictions of the energy loss of the $\tau$ at~10 PeV by different theoretical models
was found to be $+$50\% \cite{Armesto08}.

Finally, as we did not use typical values for the Earth's crust,
we considered the uncertainty in crust density itself.
Because the incident direction of $\nu_{\tau}$ events in the effective FOV
is almost horizontal, we adopted the constant value of 2.9~g/cm$^3$ \cite{Mooney98} 
which is the density of basalt, the dominant component in the  
surface layer of the Earth's crust in the vicinity of Big Island of Hawaii.
We took $-$10\% as a conservative systematic uncertainty on the 
density, which is the difference between the average density of Earth's crust 
(2.6~g/cm$^3$)\cite{PREM} and the density of basalt.

In Ref.\cite{AshraCNeu}, we evaluated the systematic uncertainty 
for the resultant upper limit by conservatively assuming
a linear or reciprocal relationship with the uncertainties 
when we considered the involved physical processes.  
In this paper, to evaluate the relation between each
source of systematic errors and the resultant upper limit in detail, we performed a number of simple Monte Carlo
simulations using TauSim with different parameter sets. 
The lower cut on the shower energy to simulate the trigger decision in the 
simplified simulation was estimated using full Monte Carlo simulation results 
obtained with CORSIKA. Using the estimated energy cut, the energy dependent 
apertures and effective areas for 1-LC and commissioning observation 
match the full simulation within 13\% and 16\% precision, respectively. 
Effects on fluence/flux limit were evaluated assuming a typical $E^{-2}$ tau neutrino flux.
The obtained systematic errors for the resultant upper limit
are summarized in Table \ref{tab:systsum}, 
where the column labeled "Estimation method" represents the uncertainty
due to using the simple Monte Carlo simulations described above, and thus the factors 
of 1.13 and 1.16 should be multiplied to the other error sources.
\begin{table}[hbt!]
\begin{center}
\begin{tabular}{cccc}
\hline
Syst. Error Source	& Source Error & 1-LC (diffuse) & Commissioning (GRB081203A) \\ 
\hline
$\nu$ Cross Section	& $-35$\%  & $32$\%   & $37$\% \\ 
$\tau$ Energy Loss	& $+50$\%  & $14$\%   & $14$\% \\ 
Rock Density		& $-10$\%  & $5$\%    & $4$\%  \\ 
Geodetic Model		& ---      & $13$\%   & $2$\%  \\ 
Estimation Method	& ---      & 13\%     & 16\%   \\ 
\hline
\end{tabular}
\end{center}
\caption{Summary table of systematic errors for the resultant upper limit. 
	The column labeled "Estimation method" represents the uncertainty
	due to the simpler limit calculation method which omit the CORSIKA 
	shower generation and the detailed trigger decision.}
\label{tab:systsum}
\end{table}

As to the geodetic model, we used detailed topographical data of the area around Mauna Kea 
and our own measurements of the mountain shape from the Ashra site to evaluate 
the effects on the resultant limit. 
Although a simple 5-sided pyramid was defined 
in TauSim to calculate tau lepton fluxes emerging from the mountain, 
the residuals from using the detailed topography were checked and the effect 
on the limit was estimated by considering a smaller and a larger 5-sided pyramid, 
covering the mountain shape by 84\% and being 84\% covered by it, respectively.  
It is negligible for observation of GRB081203A in the commissioning 
observation described in Ref.\cite{AshraCNeu}.
In 1-LC observation, the effect is slightly larger and was estimated to be
13\% for the limit on diffuse neutrino flux. The effects caused by the different density 
of the ocean compared to rock were also included in the study. 

Deviation from the linear relation in the propagation of the cross-section systematic error 
is due to the shielding of tau neutrinos by the Earth. We confirmed that the smaller cross section
results in higher sensitivity in the higher energy region at lower elevation where the chord length 
of skimmed Earth is long enough to start shielding incoming neutrinos.
Only a weak relation between tau energy losses in the rock and the sensitivity was observed at lower energy
due to the $\tau$s predominantly decaying before any significant energy loss.  
Tau energy losses
become more and more important at higher energies, and thus lower energy loss results in
higher sensitivity.  
However, this relation reverses at  $E_{\nu} \geq 1$~EeV where the chord 
length is short enough 
for not the whole $\tau$ energy being absorbed, because
$\tau$s of lower energy appearing from the Earth have a higher probability to decay in front of 
the light collector.  
In this situation, higher energy loss results in higher sensitivity. 
Due to the combination of these facts, the systematic uncertainty in $\tau$ energy losses is not 
a major factor in the sensitivity estimation.
Although uncertainty on the rock density affects both the neutrino interaction and $\tau$ energy losses,
the resultant uncertainty on the sensitivity is small because the effects basically cancel 
each other out.

\subsection{Angular resolution}\label{sec:angreso}
As discussed in Section \ref{sec:tauang}, 
Cherenkov $\tau$ shower of $E>1$~PeV  preserves the arrival direction of  
the parent $\nu_{\tau}$ to within 0.1$^\circ$ accuracy. 
This means that the detector's ability to reconstruct the arrival direction precisely
results in the identification of the VHE neutrino sources and  
leads to the realization of ``multi particle astronomy''.
Owing to its high-resolution imaging capability, the Ashra-1 detector has a huge potential to
improve the reconstruction of the arrival direction of $\nu_{\tau}$ induced air showers. 
In this section, 
shower reconstruction with likelihood analysis will be discussed,
where simulated air-shower images are generated with faster Gaisser-Hillas
and NKG parametrization \cite{TANeu}, which are well established \cite{GH-HiRes, NKG-KASKADE}. 

In this study, we adopted the following geometrical event reconstruction parameters: 
\begin{description}
\item[$(n_x, n_y)$:] 	Shower axis direction in the obtained fine image;
\item[$(X', Y')$:]	Intersection of mountain surface with the shower axis (tau emerging point) 
			projected into the obtained fine image.
\item[$E$:]		Shower energy.
\end{description}
$(n_x, n_y)$ and $(X', Y')$ determines the geometrical relationship between the detector 
and the shower axis, and therefore the impact parameter ($R_{\rm P}$).
Easy and direct comparison between real and simulated data was attained
by adopting the positions in the obtained image as the parameters
for event reconstruction. 
The positions in the obtained image correspond to the directions in the 
light collector's FOV.
For investigation of the reconstruction capability, we simulated a 10 PeV proton
shower incoming at a zenith angle of 65$^\circ$. For this type of event, 
the definition of $(X', Y')$ has to be different. 
We took the intersection of 25~km height with the shower axis
projected into the obtained image as $(X', Y')$.

\begin{figure*}[hbtp]
   \begin{center}
 \begin{tabular}{cc}
  \begin{minipage}{0.48\hsize}
   \begin{center}
    \includegraphics[width=0.98\hsize]{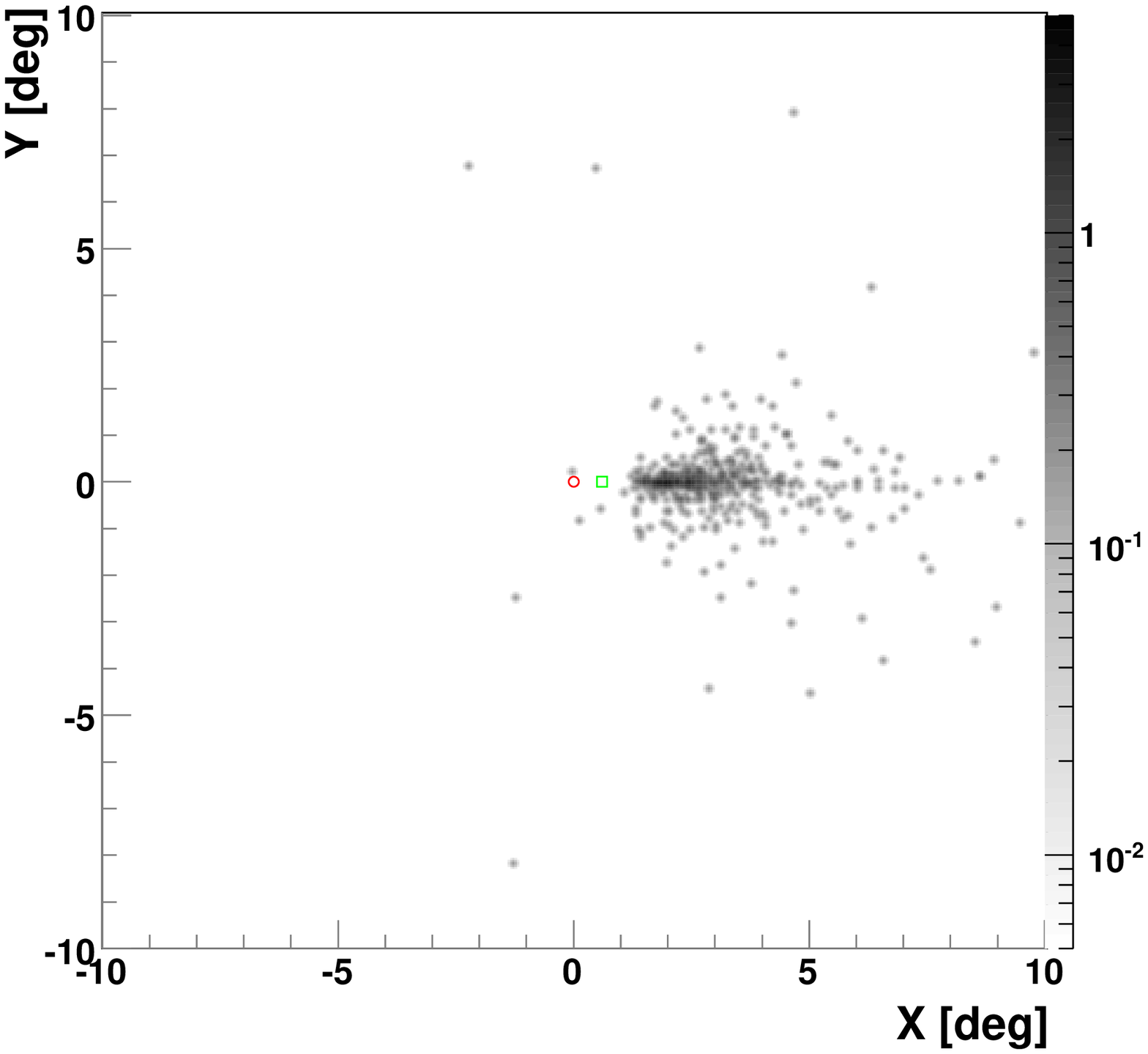}
   \end{center}
  \end{minipage} &
  \begin{minipage}{0.48\hsize}
   \begin{center}
    \includegraphics[width=0.98\hsize]{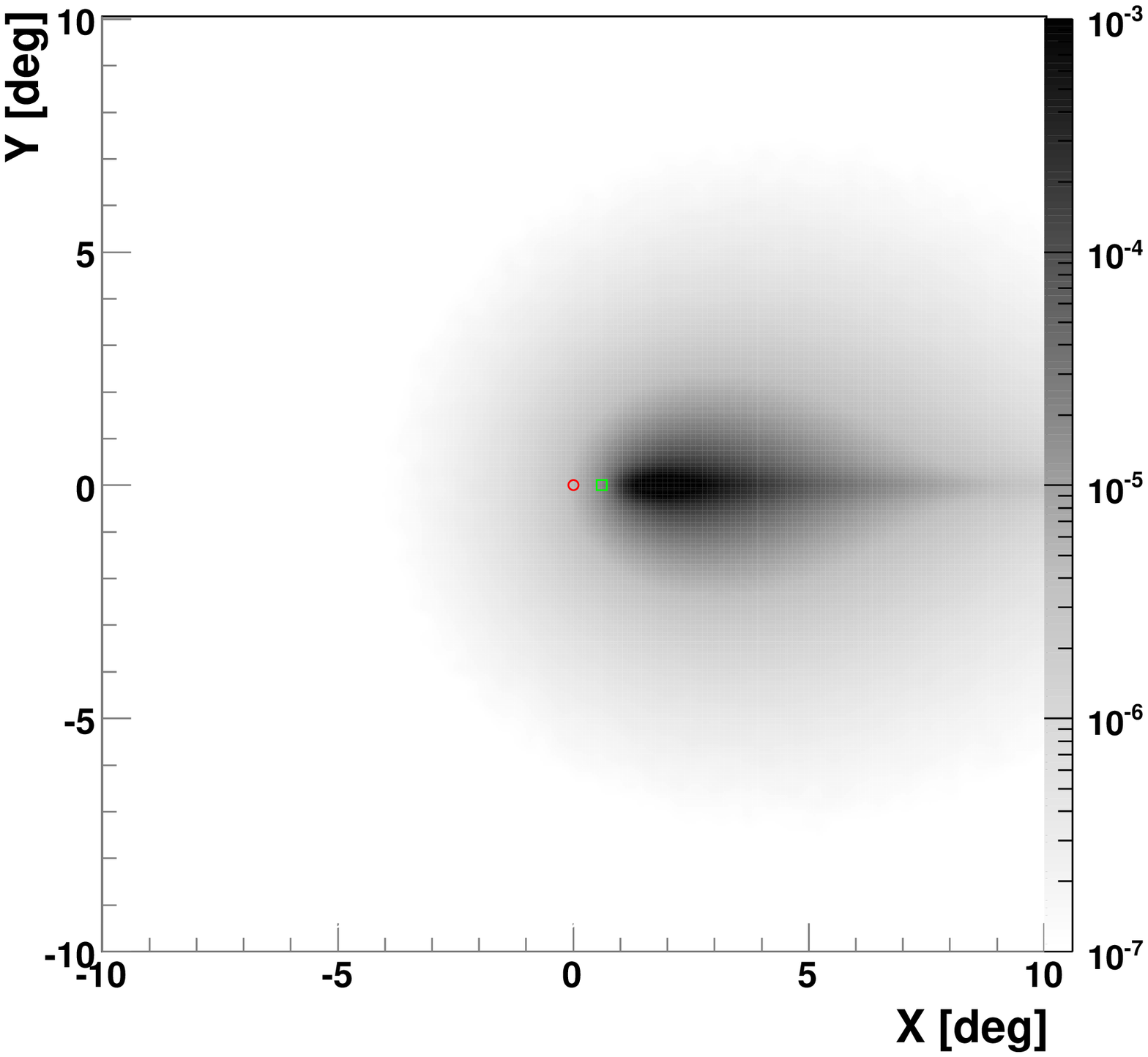}
   \end{center}
  \end{minipage} \\
 \end{tabular}
 \caption{({\it Left}) An example of Cherenkov shower image for event reconstruction ("dummy real data"),
	({\it Right}) Probability density distribution. 
	Red open circles represents $(n_x, n_y)$ and  green open square represents $(X', Y')$.
	Both images are obtained using the simulated air-showers generated using Gaisser-Hillas and NKG.
	}
 \label{fig:fluos}
 \end{center}
\end{figure*}

The following geometrical reconstruction parameters were taken 
as a typical shower example to study the reconstruction accuracy:
\begin{description}
\item[$(n_x, n_y)$] 	= (0.0, 0.0)~[deg];
\item[$(X', Y')$]	= (0.6, 0.0)~[deg];
\item[$E$]		= 10~[PeV]. 
\end{description}
This parameter set corresponds to $R_{\rm P}=$540~m. 
The left panel of Figure~\ref{fig:fluos} contains an example of the 
shower image generated with this parameter set, 
and it serves as a 
"dummy real data" to study reconstruction accuracy.
The right panel of Figure~\ref{fig:fluos} shows 
the probability density distribution obtained by averaging 10$^6$
events generated with the same parameter set.
To generate the shower images, the longitudinal and lateral development 
of air shower was calculated using Gaisser-Hillas and NKG, respectively.
The direction of Cherenkov photon was calculated by using the parametrization
described in Ref. \cite{FlysEye}, where normalization was adjusted to reproduce the $R_{\rm P}$
dependence of the detected number of photoelectrons ($N_{\rm pe}$) obtained with CORSIKA. 
We confirmed that the $R_{\rm P}$ dependence is reproduced within $\pm$15\%  
within the $R_{\rm P}$ range we used in this study. 
In this simulation, fluctuation due to the first interaction point was taken into account, 
but air-shower fluctuation due to hadron interaction was not.
This effect was accounted for 
in Section \ref{sec:tauang} (0.1$^\circ$ at 1~PeV).
Although it is preferable to include hadron air-shower fluctuation using CORSIKA,
generating a sufficient number of events requires 
an amount of CPU power unavailable during the writing of this paper.

Using the number of photoelectrons in pixel $i$ ($N_{\rm pe}^i$)
and the photon detection probability at the pixel ($p_i$), 
the log likelihood multiplied by $-2$, which is compatible with chi-square,
can be written as follows:
\begin{eqnarray}
L & = & -2 \sum_{i} N_{\rm pe}^i \log (p_i) . 
\end{eqnarray}
Note that the $N_{\rm pe}^{i}$ in each pixel was not discretized,
as the photoelectric image pipeline distributes the corresponding charge 
because of its finite resolution. 
To estimate errors on the reconstruction parameters, it is necessary to calculate the
error matrix for the likelihood function defined above. 
The error matrix is given as the inverse of the second order partial differential matrix of 
the likelihood function as follows: 
\begin{eqnarray}
L & = & L(a_{i}^{*},a_{j}^{*}) 
	+ \frac{1}{2} (a_i - a_i^*)^2 \frac{\partial^2 L}{\partial a_i^2}
	+ \frac{1}{2} (a_j - a_j^*)^2 \frac{\partial^2 L}{\partial a_j^2}
	+ (a_i-a_i^*)(a_j-a_j^*)\frac{\partial^2 L}{\partial a_i \partial a_j} \\
H_{ij} & = & 2 M_{ij} = \frac{\partial^2 L}{\partial a_i \partial a_j} \\ 
R & = & M^{-1} \label{eq:err},
\end{eqnarray}
where $a_i$ represents the parameter and  $a_i^*$ represents the parameter value 
corresponding to the local minimum. 
$R$ represents the error matrix, and $M$ represents the 
second order partial differential matrix.
$M$ can be calculated by using the neighborhood values of $L$ around the local minimum. 
\begin{equation} 
\frac{\partial^2 L}{\partial a_i \partial a_j} = \left( L - L(a_{i}^{*},a_{j}^{*}) 
	- \frac{1}{2} (a_i - a_i^*)^2 \frac{\partial^2 L}{\partial a_i^2}
	- \frac{1}{2} (a_j - a_j^*)^2 \frac{\partial^2 L}{\partial a_j^2} \right)
\times \frac{1}{(a_i-a_i^*)(a_j-a_j^*)}
\end{equation}
The inverse of $M$ gives $R$ and, 
once the error matrix is obtained, the diagonal elements correspond to
the parameter errors including correlation between parameters. 
Elements of the correlation matrix are given by the following equation. 
\begin{equation}
C_{ij} = \frac{R_{ij}}{\sqrt{R_{ii} R_{jj}}}
\end{equation}
Using the "dummy real data" events generated by the Monte Carlo simulation, 
the correlation between reconstruction parameter was evaluated using the 
inverse matrix calculations described above. 
For simplicity, the energy of the shower was fixed for this calculation. 
The following example of an obtained correlation matrix shows that there are strong
correlations between $n_x$ and $X'$. 
\begin{eqnarray*}
\begin{array}{ccccc}
    & n_x  & n_y  & X'   & Y' \\
n_x & 1.00 & 0.04 & 0.95 & 0.03 \\ 
n_y & 0.04 & 1.00 & 0.04 & 0.78 \\
X'  & 0.95 & 0.04 & 1.00 & 0.02 \\
Y'  & 0.03 & 0.78 & 0.02 & 1.00 \\
\end{array}
\end{eqnarray*}

Considering the fact that $X'$ determines $R_{\rm P}$,
strong correlation between $n_x$ and $R_{\rm P}$ results in 
a worse joint resolution.
This is a well-known problem connected with the mono detection of Cherenkov shower
and is greatly improved in case of stereo detection with multiple light collectors.  
In order to estimate the angular resolution of the reconstruction, 
therefore, we scanned $n_x$ and $X'$ space with 0.05$^\circ$ grid to 
search for the minimum of $L$ while fixing $n_{y}$ and $Y'$.
To keep $N_{\rm pe}$ constant over different $R_{\rm P}$ data sets, 
the energy of the shower was adjusted. Since $N_{\rm pe}$ and $R_{\rm P}$ are 
inversely correlated, 
parameter sets with larger $R_{\rm P}$ are given larger shower energy. 
200 events of "dummy real data" were generated using the same parameter set,
but with different first interaction points and different random seeds. 
Since the maximum likelihood method uses the slope of probability density 
distribution, it is important to include the night sky background (NSB) photoelectrons.
NSB photoelectrons were added as a probability per pixel in probability density 
distribution.
To each ¡Édummy real data¡É event, photonelectrons randomly distributed according to
this probability were generated and added as well. 
We estimated the region for which the likelihood is calculated by using each "dummy real data" 
by defining cut positions which correspond to 5\% of the peak in the projections 
of major and minor axes. In the major and minor axis projection, we used asymmetric and symmetric 
Gaussian to fit their shape, respectively, and calculated the 5\%-peak positions to reduce the 
effects of statistical fluctuation toward the tails of the distributions.

After selecting ($n_x, X'$) which gave minimum $L$ in each "dummy real data", the distribution
of residuals between the reconstructed and true directions ($\Delta n_{x}$; defined as 
reconstructed minus true positions in the obtained image) was obtained.
\begin{figure*}[bt!]
\begin{center}
\begin{tabular}{cc}
 \begin{minipage}{0.46\hsize}
  \begin{center}
    \includegraphics[width=0.98\hsize]{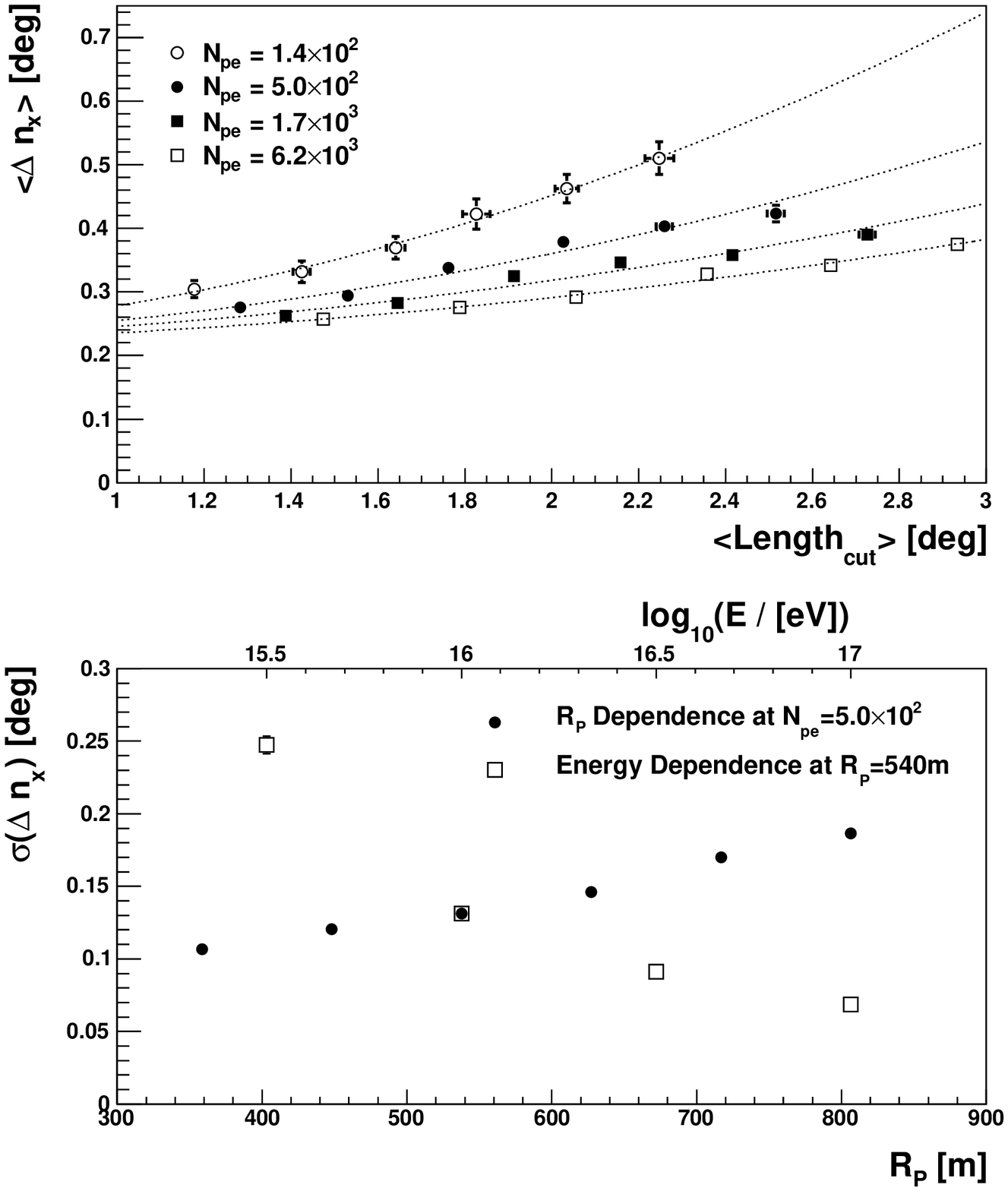}
  \end{center}
 \end{minipage} &
 \begin{minipage}{0.42\hsize}
  \begin{center}
   \includegraphics[width=0.98\hsize]{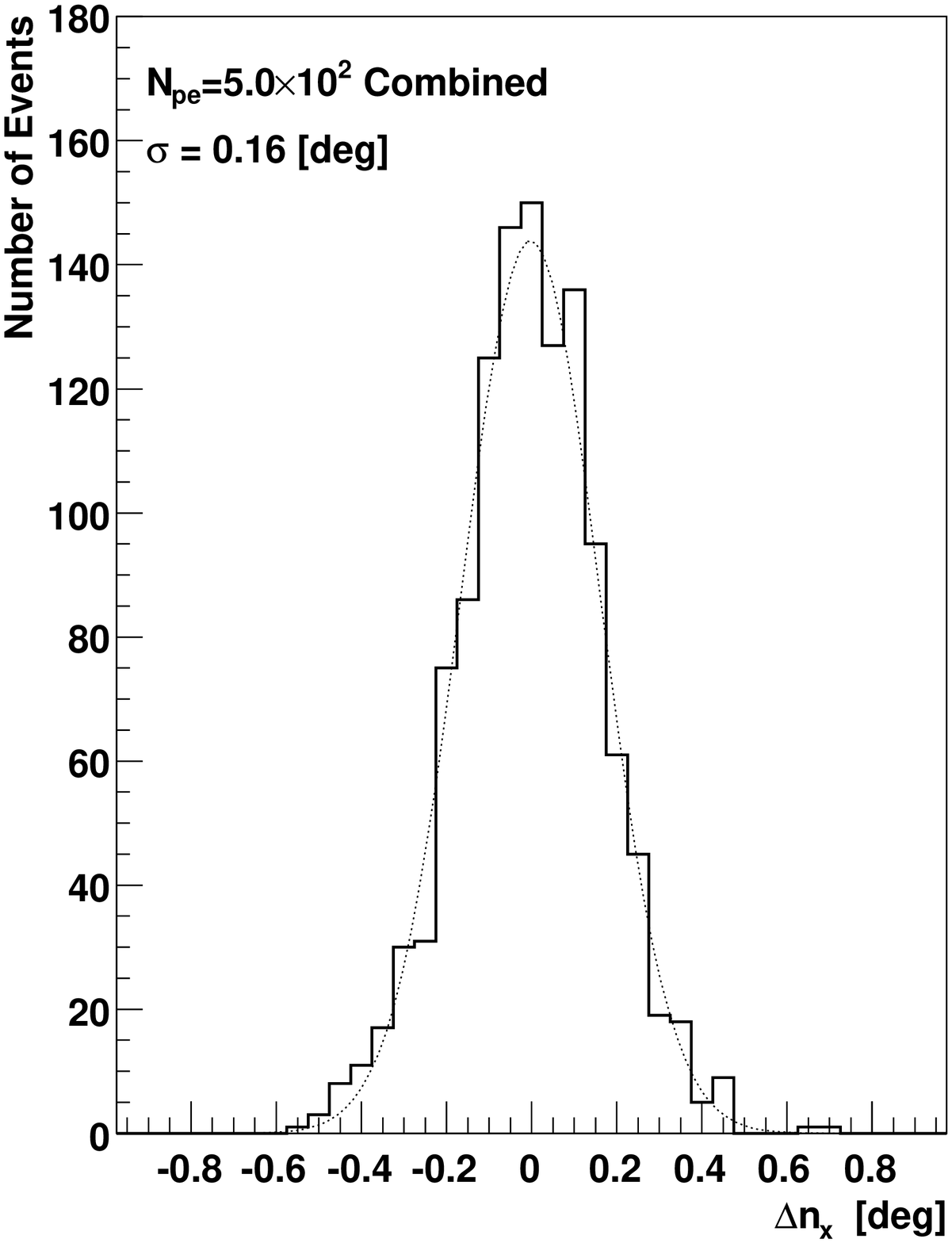}
  \end{center}
 \end{minipage} \\
 \end{tabular}
\end{center}
 \caption{
	({\it Upper Left}) The dotted curves represent correction function for $\Delta n_{x}$ offset
	($F_{C}(N_{\rm pe}, Length_{\rm cut})$). Open circles, closed circles, 
	closed squares, and open squares represents the relation between $<\Delta n_{x}>$ 
	and $<Length_{\rm cut}>$ for $N_{\rm pe}=$140, 500, 1700, and 6200, respectively.
	({\it Bottom Left}) The $R_{\rm P}$ dependence (closed circles) and energy dependence 
	(open squares) of the angular resolution along 	the major axis ($\sigma (\Delta n_{x})$).
	Energy dependence is shown at the fixed $R_{\rm P}$ of 540m while the $R_{\rm P}$ dependence
	is shown at fixed $N_{\rm pe}$ of 500.
	({\it Right}) Residuals between the reconstructed and true directions ($\Delta n_{x}$) 
	after correcting $\Delta n_x$ offset. The distribution includes all the parameter sets
	of ($n_x, X') = (0.0,0.4), (0.0,0.5), (0.0,0.6), (0.0,0.7), (0.0,0.8), (0.0,0.9)$~[deg] for $N_{\rm pe}=500$
	(see the text).
	}
 \label{fig:angreso}
\end{figure*}
We found that the distribution has an offset toward positive side, meaning the smaller $R_{\rm P}$, 
corresponding to smaller $X'-n_{x}$, tends to have minimum $L$.
This was mainly due to the limited range of the likelihood calculation forced by the existence of 
NSB photoelectrons.
In order to correct for this effect, 
application of a correction as a function of observable quantities was necessary. 
The observable quantities required for the correction are 
$N_{\rm pe}$ and an estimator of $R_{\rm P}$. 
We used $Length_{\rm cut}$ as the estimator of $R_{\rm P}$, where
$Length_{\rm cut}$ is defined as the 5\%-peak cut length along the major axis.
The upper left panel of Fig.\ref{fig:angreso} shows the resultant correction function 
for $\Delta n_{x}$ offset ($F_{C}(N_{\rm pe}, Length_{\rm cut})$) as dotted curves.
Open circles, closed circles, closed squares, and open squares represents the relation 
between $<\Delta n_{x}>$ and $<Length_{\rm cut}>$ for $N_{\rm pe}=$140, 500, 1700, and 6200, 
respectively, where $<\Delta n_{x}>$ and $<Length_{\rm cut}>$ denotes the averaged 
value of $\Delta n_{x}$ and $Length_{\rm cut}$ for each parameter set, respectively. 
For each $N_{\rm pe}$ data set, $X'$ was scanned from 0.4$^\circ$ to 0.9$^\circ$ in step of 0.1$^\circ$,
corresponding to the points from left to right. 
Each $N_{\rm pe}$ data set was used to obtain $F_{C}(N_{\rm pe}, Length_{\rm cut})$
with the fit to second order polynomial. 

The $R_{\rm P}$ dependence on the angular resolution of $n_x$ ($\sigma ( \Delta n_x )$) is shown in the bottom left panel
in Fig.~\ref{fig:angreso} as closed circles, together with the energy dependence on $\sigma ( \Delta n_x )$ as open squares, 
where $\sigma ( \Delta n_x )$ corresponds to the root mean square of the $\Delta n_x$ distribution.
Energy dependence is shown at the fixed $R_{\rm P}$ of 540m while the $R_{\rm P}$ dependence
is shown at fixed $N_{\rm pe}$ of 500.
Slight worsening of the angular resolution is visible as $R_{\rm P}$ increases, where 
the data corresponds to $(n_x,X')=(0.0,0.4), (0.0,0.5), (0.0,0.6), (0.0,0.7), (0.0,0.8), (0.0,0.9)$,
respectively, from left to right.
On the other hand, clear worsening of the angular resolution proportional to square root of $N_{\rm pe}$
is observed in energy dependence, while a resolution better than 0.1$^\circ$ is obtained with  
sufficient $N_{\rm pe}$ at higher energies.

After applying the correction function in an event-by-event basis using the calculated $Length_{\rm cut}$
and $N_{\rm pe}$ for each "dummy real data" sample, the $\Delta n_x$ distribution from all parameter sets 
was obtained as shown in the right panel of Fig.\ref{fig:angreso}.
The resultant angular resolution was estimated to be 0.16$^\circ$.
Considering the fact that $N_{\rm pe}$ for the parameter sets $N_{\rm pe}=500$ 
and the expected $N_{\rm pe}$ from tau 
neutrino signal as discussed in Section~\ref{sec:sen}, the geometrical parameters used here
would be appropriate to estimate the background contamination since they correspond to the events
with less statistical power.
This analysis outlined the fact that the high-resolution Cherenkov 
imaging would lead to the high accuracy reconstruction of the arrival direction,
even with the mono detection of Cherenkov showers.

\subsection{Background}\label{bgd}
In this section, we evaluate the background events due to air showers.
Background events due to the detector itself are discussed in Ref. \cite{AshraCNeu}. 
Air-shower background candidates are normal cosmic rays, muons, muon neutrinos, 
$\tau$s, and $\nu_{\tau}$s. 
From simple flux calculations, it was shown that 
the neutrino components through mountain, prompt $\tau$ components and muons
are negligible \cite{Noda09, Zas93, Enberg08, Martin03, Athar03}.
Thus, the large zenith angle component of normal cosmic rays was considered 
as the dominant background component in this study. 
To simulate this background, 
CORSIKA was used in the same way as with the $\nu_{\tau}$ simulation.
To consider the atmospheric depth correctly, "CURVED EARTH" option was selected.
The IACT package was used to simulate incoming Cherenkov photons.
Protons were selected as the parent particle and a thinning parameter of 10$^{-5}$ 
was applied.
To investigate the zenith angle and the energy dependence of the background flux,
a difference in thinning level between this simulation and the $\nu_{\tau}$ 
simulation was unavoidable. 
The bunch size distribution of the generated showers was studied in this case, too.
We found that 74\% (93\%) of them had 20 (5) bunches or more 
when detected number of photoelectrons was 40 or more.
The applied thinning level of 10$^{-5}$ was acceptable in terms of 
background estimation precision. 

At first, to estimate the background rate during the commissioning observation,
the trigger rate for trigger-pixel arrangement adopted in the commissioning observation was
calculated.  
As we could not simulate the largest zenith angle ($\theta_{\rm zen}$) of 
$\theta_{\rm zen} > 88^\circ$ in the combination of "CURVED EARTH" and IACT options,
trigger pixels were offset by two degrees toward higher elevation 
to estimate the trigger rate for $88^\circ < \theta_{\rm zen} < 90^\circ$.
To account for the trigger rate decrease due to thicker atmospheric depth,
the result was further corrected by the ratio of expected events
between $84^\circ < \theta_{\rm zen} < 86^\circ$ and
$86^\circ < \theta_{\rm zen} < 88^\circ$ in the event that the entire FOV was
covered by the trigger pixel. 
As a result, the number of expected background events ($N_{\rm CR}$) 
due to normal cosmic rays during commissioning observation 
of 197.1~hr was estimated to be:
\[ N_{\rm CR} = 1.3 \times 10^{-4}, \]
which is small enough to be neglected.  
Note that the above discussion did not use an event selection to discard cosmic rays
using the reconstructed arrival direction information.

Next, the cosmic-ray background in 1-LC observation with final configuration was evaluated.
With the maximum weather efficiency of 100\%, 1750~hr of observation time is expected 
in one year.
Assuming a trigger pixel threshold of 20 photoelectrons which is the same as for our 
sensitivity estimation, the cosmic-ray shower event rate emerging from the
sky region near the mountain edge was found to be 8.2$\times 10^{-2}$, 0.55, 4.3, 39 (per year)
within 0.1$^\circ$, 0.3$^\circ$, 1.0$^\circ$, 3.0$^\circ$ from the mountain edge,
respectively.
Assuming perfect reconstruction of the arrival direction, 
a background-free result is achieved by requiring that the reconstructed direction  
of the shower axis points back toward the mountain or the Earth  
because they completely absorb any cosmic-ray secondary particles. 
In practice, the background-free result will be achieved by requiring a mountain-edge cut 
in which the reconstructed arrival direction is inside the mountain edge by  
$\theta_{\rm cut}$, where $\theta_{\rm cut}$ is an angle dependent on the  
reconstruction accuracy.
The resultant yearly background rate ($N_{\rm CR}^{\rm 1 LC}$) is calculated as follows:
\begin{eqnarray}
 N_{\rm CR}^{\rm 1 LC} (\theta_{\rm cut}) & = &
	\int_{0}^{\theta_{\rm max}} n_{\rm CR}(\theta) 
	\int_{0}^{\pi} p_{\rm axis}(\phi) 
	\int^{-(\theta + \theta_{\rm cut}) / | \sin \phi |}_{-\infty} p_{\rm rec}(\theta') \, d \theta' \, d \phi \, d \theta 
	\nonumber \\
	& + &
	\int_{0}^{\theta_{\rm max}} n_{\rm CR}(\theta) 
	\int_{-\pi}^{0} p_{\rm axis}(\phi) 
	\int^{\infty}_{(\theta + \theta_{\rm cut}) / | \sin \phi |} p_{\rm rec}(\theta') \, d \theta' \, d \phi \, d \theta  
\end{eqnarray}
where $\theta$ is defined as the vertical angle from the mountain edge (positive toward the sky region), 
$\theta_{\rm max}$ represents the maximum angle from the mountain edge in the FOV, 
$\phi$ represents the angle between major axis of the shower image and the horizontal axis,
$p_{\rm axis}(\phi)$ denotes the probability to obtain the major axis at an angle $\phi$,
$p_{\rm rec}(\theta)$ denotes the misreconstruction probability distribution to obtain the 
angular deviation from the true value by $\theta$ along the major axis,
and $n_{\rm CR}(\theta)$ denotes the background rate per radian.  
In this calculation, the deviation along the minor axis is neglected. 
An example of $p_{\rm rec}(\theta)$ is shown in the right panel of Fig.\ref{fig:angreso},
where $\Delta n_{x}$ corresponds to $\theta$.
$p_{\rm axis}(\phi)$ is normalized as
\begin{equation}
 \int_{-\pi}^{\pi} p_{\rm axis}(\phi) d \phi = 1 \,.
\end{equation}
Using this notation, for example, the expected background rate emerging from 
the sky region within 1.0$^\circ$ from the mountain edge can be expressed as follows:
\begin{equation}
 \int_{0}^{1.0^\circ} n_{\rm CR}(\theta) d \theta = 0.55 \,.
\end{equation}
$\theta_{\rm cut}$ will be determined so that the resultant $N_{\rm CR}^{\rm 1 LC} (\theta_{\rm cut})$ is small enough.
Although it requires a more detailed procedure to apply this background rejection scheme for real data analysis,
the discussion in this section and the previous section clearly illustrate its working principle. 

It is important to evaluate the effective aperture dependence on the mountain-edge
cut. By using the neutrino simulation data set, the effective aperture was 
estimated to be 90\%, 80\% and 38\% of that without the cut
for the mountain-edge cuts of 0.3$^\circ$, 1.0$^\circ$ and 3.0$^\circ$, respectively. 
The right panel of Figure~\ref{fig:aper} shows the energy dependence of effective
apertures with the mountain-edge cuts by open symbols. 
Blue open squares, blue open triangles, and blue open diamonds represent the
effects of the mountain-edge cuts of 0.3$^\circ$, 1.0$^\circ$, and 3.0$^\circ$, 
respectively.
The effective aperture with the 3.0$^\circ$ mountain-edge cut was shown to 
estimate the $\theta_{\rm cut}$ dependence of the effective aperture 
although 3.0$^\circ$ is too large compared with the Ashra-1 angular resolution. 
The large decrease in the effective aperture at higher energy in the 3.0$^\circ$ 
mountain-edge cut is due to the fact that higher energy neutrinos could only be  
emerging from the mountain edge because of smaller interaction length. 
Since larger mountain-edge cut directly affects the effective aperture of the detector,
the reconstruction accuracy of the arrival direction is important to realize the 
background-free observation while keeping a high sensitivity. 
In addition, it is also very important to positively identify the neutrino shower  
events.  
The high-resolution imaging capability would be a key feature 
in the detection of VHE neutrinos for the first time.

\section{Conclusion}\label{sec:concl}
In the observation of VHE neutrinos from high energy transient objects,
highly precise arrival direction determination
is essential to the following points: 
(1) Clear identification of neutrinos by distinguishing their appearance from the Earth;
(2) Identification of VHE neutrino point sources.
We studied the physical processes of Earth-skimming $\nu_{\tau}$s in detail,
and found that 
Cherenkov $\tau$ showers of more than 1~PeV energy  preserve the arrival direction of the 
parent $\nu_{\tau}$ 
with an accuracy better than 0.1$^\circ$. 
The Ashra detector uses newly developed light collectors 
which realize both a 42 degree-diameter FOV and arcminute resolution. 
It is capable of capturing the high-resolution images of Cherenkov
emission with unprecedented precision. 
Based on fine imaging data,
the use of maximum likelihood analysis would lead to 0.1--0.2$^\circ$ accuracy in 
the determination of the arrival direction of Cherenkov showers even with 
monoscopic detection by only one light collector. 
The 
``Cherenkov $\tau$ shower ES method'' which detects the Cherenkov photons from $\tau$ air shower
has the highest point source sensitivity in the 10--100PeV range.
These results indicate that the
Ashra-1 detector 
is a unique probe of VHE neutrinos and their accelerators.
The first physics results using this method were published in Ref. \cite{AshraCNeu}.

\section*{Acknowledgment}
We thank H.~Motz, S.~Ogawa, P.~Binder and J.~Goldman for useful discussion. 
The Ashra-1 project is supported by  
the Coordination Fund for Promoting Science and Technology (157-20004100) 
and by Grant-in-Aid for Scientific Research (19340055, 19403004, 16403001, 23684013) 
from the Ministry of Education, Culture, Sports, Science and Technology
in Japan. 


\begin{thebibliography}{10}
\expandafter\ifx\csname url\endcsname\relax
  \def\url#1{\texttt{#1}}\fi
\expandafter\ifx\csname urlprefix\endcsname\relax\def\urlprefix{URL }\fi
\expandafter\ifx\csname href\endcsname\relax
  \def\href#1#2{#2} \def\path#1{#1}\fi

\bibitem{WB95}
E.~Waxman, J.~Bahcall, \prl 75 (1995) 386.

\bibitem{Vietri95}
M.~Vietri, \apj 453 (1995) 883.

\bibitem{Milgrom95}
M.~Milgrom, V.~Usov, \apj 449 (1995) L37.

\bibitem{BeppoSax97}
E.~Costa, et~al., Nature 387~(6635) (1997) 783--785.

\bibitem{Rees-Meszaros92}
M.~J. Rees, P.~Meszaros, MNRAS 258 (1992) 41.

\bibitem{Sari-Piran95}
R.~Sari, T.~Piran, ApJ Lett. 455 (1995) L143.

\bibitem{Piran99}
T.~Piran, Phys. Rep. 314 (1999) 575.

\bibitem{Meszaros06}
P.~M\'es\'zaros, Rep. Prog. Phys. 69 (2006) 2259, and references therein.

\bibitem{Gehrels04}
N.~Gehrels, et~al., \apj 611 (2004) 1005.

\bibitem{Abdo09a}
A.~A. Abdo, et~al., Science 323 (2009) 1688.

\bibitem{Sasaki00}
M.~Sasaki, Proc. of ICRR2000 Satellite Symposium: Workshop of Comprehensive
  Study of the High Energy Universe.

\bibitem{Barwick00}
S.~Barwick, Physica Scripta. T85 (2000) 106.

\bibitem{IceCubeGRBs}
R.~Abbasi, et~al., \apj 710 (2010) 346.

\bibitem{IceCubeGRB}
R.~Abbasi, et~al., \apj 701 (2009) 1721.

\bibitem{Domokos98}
G.~Domokos, S.~Kovesi-Domokos, AIP Conf. Proc. 433 (1998) 390.

\bibitem{Letessier00}
A.~Letessier-Selvon, AIP Conf. Proc. 566 (2000) 157.

\bibitem{Athar00}
H.~Athar, G.~Parente, E.~Zas, \prd 62 (2000) 093010.

\bibitem{Fargion02}
D.~Fargion, \apj 570 (2002) 909.

\bibitem{Feng02}
J.~Feng, P.~Fisher, F.~Wilczek, T.~Yu, \prl 88 (2002) 161102.

\bibitem{Abraham08}
J.~Abraham, et~al., \prl 100 (2008) 211101.

\bibitem{RMW04}
S.~Razzaque, P.~M\'es\'zaros, E.~Waxman, \prd 69 (2004) 023001.

\bibitem{IAUC6895}
C.~Tinney, et~al., IAU Circular, 6895.

\bibitem{GCN4792}
N.~Mirabal, et~al., GCN Circular, 4792.

\bibitem{Sasaki08}
M.~Sasaki, J. Phys. Soc. Jpn. 77SB~(Supplement B) (2008) 83.

\bibitem{PLI11}
Y.~Asaoka, M.~Sasaki, Nucl. Instrum. Methods Phys. Res. A 647 (2011) 34.

\bibitem{AshraCNeu}
Y.~Aita, et~al., \apjl 736 (2011) L12.

\bibitem{GCN8632}
Y.~Aita, et~al., GCN Circular, 8632.

\bibitem{GCN11291}
Y.~Asaoka, et~al., GCN Circular, 11291.

\bibitem{MaunaKea}
E.~Wolfe, W.~Wise, G.~Dalrymple, U.S. Geological Survey Professional Paper 1557
  (1997) 129.

\bibitem{PYTHIA6154}
T.~Sj{\" o}strand, et~al., Comput. Phys. Commun. 135 (2001) 238.

\bibitem{Geant4}
S.~Agostinelli, et~al., Nucl. Instrum. Methods A 506 (2003) 250.

\bibitem{Dutta2001}
S.~Iyer~Dutta, et~al., \prd 63 (2001) 094020.

\bibitem{ALLM}
H.~Abramowicz, A.~Levy, arXiv:hep-ph/9712415v2.

\bibitem{DiffCS}
B.~Badelek, J.~Kwiecinski, Rev. Mod. Phys. 68 (1996) 445.

\bibitem{TAUOLA24}
S.~Jadach, et~al., Comput. Phys. Commun. 76 (1993) 361.

\bibitem{NuTel}
G.~W. Hou, M.~Huang, astro-ph/0204145.

\bibitem{TANeu}
M.~Sasaki, Y.~Asaoka, M.~Jobashi, Astropart. Phys. 19 (2003) 37.

\bibitem{Noda09}
Y.~Aita, et~al., in: 31th Intl. Cosmic Ray Conf. (Lodz), ID0313, 2009.

\bibitem{Gandhi98}
R.~Gandhi, et~al., \prd 58 (1998) 093009.

\bibitem{Gandhi96}
R.~Gandhi, et~al., Astropart. Phys. 5 (1996) 81.

\bibitem{Tseng03}
J.~Tseng, et~al., \prd 68 (2003) 063003.

\bibitem{PREM}
A.~M. Dziewonski, D.~L. Anderson, Physics of the Earth and Planetary Interiors
  25 (1981) 297.

\bibitem{CORSIKA}
D.~Heck, et~al., Report FZKA 6019.

\bibitem{QGSJET01}
N.~Kalmykov, S.~Ostapchenko, A.~Pavlov, Nucl. Phys. B (Proc. Suppl.) 52B (1997)
  17.

\bibitem{Anchordoqui06}
L.~Anchordoqui, A.~Cooper-Sarkar, D.~Hooper, S.~Sarkar, \prd 74~(4) (2006)
  43008.

\bibitem{Kutak03}
K.~Kutak, J.~Kwiecinski, Eur. Phys. J. C 29 (2003) 521.

\bibitem{Henley06}
E.~Henley, J.~Jalilian-Marian, \prd 73~(9) (2006) 094004.

\bibitem{Sarkar08}
A.~Cooper-Sarkar, S.~Sarkar, JHEP 01 (2008) 075.

\bibitem{Berger08}
E.~Berger, et~al., \prd 77 (2008) 053007.

\bibitem{AugerES09}
J.~Abraham, et~al., \prd 79 (2009) 102001.

\bibitem{Kharzeev04}
D.~Kharzeev, Y.~Kovchegov, K.~Tuchin, Phys.~Lett.~B 599 (2004) 23.

\bibitem{Bartels02}
K.~Bartels, J. Golec-Biernat, H.~Kowalski, \prd 66 (2002) 014001.

\bibitem{Armesto08}
N.~Armesto, C.~Merino, G.~Parente, E.~Zas, \prd 77~(1) (2008) 13001.

\bibitem{Mooney98}
W.~Mooney, G.~Laske, T.~Masters, J.~Geophys.~Res.~ 103~(B1) (1998) 727--747.

\bibitem{GH-HiRes}
C.~Song, et~al., Astropart. Phys. 14 (2000) 7.

\bibitem{NKG-KASKADE}
T.~Antoni, et~al., Astropart. Phys. 14 (2001) 245.

\bibitem{FlysEye}
R.~Baltrusaitis, et~al., Nucl. Instrum. Methods A 240 (1985) 410.

\bibitem{Zas93}
E.~Zas, et~al., Astropart.Phys. 1 (1993) 297.

\bibitem{Enberg08}
R.~Enberg, et~al., \prd 78 (2008) 043005.

\bibitem{Martin03}
A.~Martin, et~al., Acta Physica B 34 (2003) 3273.

\bibitem{Athar03}
H.~Athar, et~al., Astropart.Phys. 18 (2003) 581.

\end{thebibliography}
\end{document}